\documentclass[showpacs]{revtex4}
\usepackage{epsf}  

\begin{document}

\title{Extra galactic sources of high energy neutrinos\footnote{Summary of talk presented at the Nobel Symposium 129: Neutrino Physics, Enk\"oping, Sweden, 2004}}

\author{Eli Waxman}
\altaffiliation{Physics faculty, Weizmann institute of science, Rehovot 76100, Israel}

\begin{abstract}

The main goal of the construction of large volume, high energy neutrino telescopes is the detection of extra-Galactic neutrino sources. The existence of such sources is implied by observations of ultra-high energy, $\ge10^{19}$~eV, cosmic-rays (UHECRs), the origin of which is a mystery. The observed UHECR flux sets an upper bound to the extra-Galactic high energy neutrino intensity, which implies that the detector size required to detect the signal in the energy range of 1~TeV to 1~PeV is $\ge 1$~giga-ton, and much larger at higher energy. Optical Cerenkov neutrino detectors, currently being constructed under ice and water, are expected to achieve $1$~giga-ton effective volume for 1~TeV to 1~PeV neutrinos. Coherent radio Cerenkov detectors (and possibly large air-shower detectors) will provide the $\gg 1$~giga-ton effective volume required for detection at $\sim10^{19}$~eV. Detection of high energy neutrinos associated with electromagnetically identified sources will allow to identify the sources of UHECRs, will provide a unique probe of the sources, which may allow to resolve open questions related to the underlying physics of models describing these powerful accelerators, and will provide information on fundamental neutrino properties. 

\end{abstract}

\maketitle

\section{Introduction and summary}
\label{sec:intro}

The detection of MeV neutrinos from the Sun enabled direct observations of nuclear reactions in the core of the Sun, as well as studies of fundamental neutrino properties \cite{Solar_nus}. Existing MeV neutrino "telescopes" are also capable of detecting neutrinos from supernova explosions in our local Galactic neighborhood, at distances $<100$~kpc, such as supernova 1987A. The detection of neutrinos emitted by SN1987A provided a direct observation of the core collapse process and constraints on neutrino properties \cite{SN_nus}. The main goal of the construction of high energy, $>1$~TeV, neutrino telescopes \cite{Nu_telescopes} is the extension of the distance accessible to neutrino astronomy to cosmological scales. 

The existence of extra-Galactic high-energy neutrino sources is implied by cosmic-ray observations. The cosmic-ray spectrum extends to energies $\sim10^{20}$~eV, and is likely dominated beyond $\sim10^{19}$~eV by extra-Galactic sources of protons \cite{CR_data_rev} (see, however, \cite{WatsonComposition}). The origin of the highest energy, $>10^{19}$~eV, cosmic rays (UHECRs) is a mystery. It has been suggested that modifications of the basic laws of physics are required in order to account for the existence of UHECRs. Such "new physics" models commonly postulate the existence of very massive particles, the decay of which produces the observed UHECRs, and generally predict large fluxes of $\sim10^{20}$~eV neutrinos (see, e.g., \cite{BhattacharjeeSigl} for review). I focus here on UHECR production models, which do not invoke modifications to the basic laws of physics. In these models it is assumed that UHECRs are charged particles, most likely protons, accelerated electromagnetically to high energy in astrophysical objects. In this case, some fraction of the protons are expected to produce pions as they escape their source by either hadronic collisions with ambient gas or photo-production with source photons, leading to electron and muon neutrino production through the decay of charged pions.

In \S\ref{sec:phenomenology} a phenomenological, model independent discussion of extra-Galactic high energy neutrino sources is presented. In \S\ref{sec:luminosity} we show that the stringent constraints, which are imposed on the properties of possible UHECR sources by the high energies observed, rule out almost all source candidates, and suggest that $\gamma$-ray bursts (GRBs) and active galactic nuclei (AGN) are the most plausible sources. We furthermore demonstrate that these constraints also imply that the UHECR sources may be detectable as point sources at $\sim1$~TeV to $\sim1$~PeV neutrino energies by km-scale (i.e. giga-ton scale) neutrino telescopes. The required large effective volume will be achieved by optical Cerenkov detectors being constructed under ice and water \cite{Nu_telescopes}. 

In \S\ref{sec:WB} we discuss the constraints imposed by cosmic-ray observations on the diffuse extra-Galactic high energy neutrino intensity produced by sources which, like GRBs and AGN jets, are optically thin for high-energy nucleons to $p\gamma$ and $pp(n)$ interactions. The upper bound (fig.~\ref{fig:WBbound}, eq.~\ref{eq:WB}), which came to be known as the Waxman-Bahcall (WB) bound \cite{WBbound}, implies that km-scale (i.e. giga-ton) neutrino telescopes are needed to detect the expected diffuse extra-Galactic flux in the energy range of $\sim1$~TeV to $\sim1$~PeV, and much larger effective volume is required at higher energy. Implications of the bound to predictions of neutrino emission from AGN (figure~\ref{fig:agn}) are briefly discussed in \S~\ref{sec:AGN}. We note, that the WB bound may be evaded by postulating the existence of sources which are optically thick for protons to $p\gamma$ or $pp(n)$ interactions \cite{WBbound,MPR99}. However, the existence of such "hidden," or "neutrino only," factories is not motivated by measurements of the cosmic-ray flux or by electromagnetic observations.

In \S\ref{sec:GZK} we discuss "GZK neutrinos." If the sources of UHECRs are extra-Galactic and the particles are indeed protons, then these particles lose energy by interacting with the cosmic microwave background photons \cite{gzk}. Protons of energy exceeding the threshold for pion production, $\sim5\times10^{19}$~eV, lose most of their energy over a time short compared to the age of the universe (the "GZK effect"). The decay of the pions generates a background of high energy neutrinos (\cite{gzk_nu}; for detailed updated discussion see \cite{ESS01} and references therein). The intensity of this background, at neutrino energies $\sim10^{19}$~eV, should be similar to the WB bound. Coherent radio Cerenkov detectors, and possibly large air-shower detectors, will provide the large effective volume required for the detection of the "GZK neutrinos" \cite{Saltzberg}. Their detection will help to determine the identity of the cosmic-ray particles and will constrain the redshift evolution of UHECR sources.

In \S\ref{sec:grb} we discuss in some detail high energy neutrino emission from GRBs. This discussion illustrates the applicability of the phenomenological, model independent arguments presented in \S\ref{sec:phenomenology} through a discussion of (a model of) a particular, plausible extra-Galactic source of UHECRs and high energy neutrinos. In \S\ref{sec:fireball} we present a brief discussion of the fireball model of GRBs, and briefly present the arguments suggesting a connection between GRBs and UHECR sources. In \S\ref{sec:1e14} we show that production of 100~TeV neutrinos in the region where GRB $\gamma$-rays are produced is a generic prediction of the fireball model. It is a direct consequence of the {\it assumptions} that energy is carried from the underlying engine, most likely a (few) solar mass black hole, as kinetic energy of protons and that $\gamma$-rays are produced by synchrotron emission of shock accelerated particles. The detection of the predicted neutrino signal will therefore provide strong support for the validity of underlying model assumptions, which is difficult to obtain using photon observations (due to the high optical depth in the vicinity of the GRB "engine"). The predicted neutrino intensity, $\approx20\%$ of the WB intensity bound, implies a detection of $\sim20$ muon induced neutrino events per yr in a km-scale neutrino detector (since these events should be correlated in time and direction with GRB $\gamma$-rays, the search for GRB neutrinos is essentially background free). Neutrinos may be produced also in other stages of fireball evolution, at energies different than 100~TeV. The production of these neutrinos is dependent on additional model assumptions. As an example, we discuss in \S\ref{sec:TeV} the production of TeV neutrinos expected in the "collapsar" scenario, where GRB progenitors are associated with the collapse of massive stars.

The discussion of GRB neutrino emission demonstrates that in addition to identifying the sources of UHECRs, high-energy neutrino telescopes can also provide a unique probe of the physics of these sources. Moreover, detection of high energy neutrinos from extra-Galactic sources may also provide information on fundamental neutrino properties \cite{WnB97}. High energy neutrinos are expected to be produced in astrophysical sources by the decay of charged pions, which lead to the production of muon and electron neutrinos. However, oscillation to $\nu_\tau$'s \cite{Osc} imply that neutrino telescopes should detect equal numbers of $\nu_\mu$'s and $\nu_\tau$'s. Up-going $\tau$'s, rather than $\mu$'s, would be a distinctive signature of such oscillations. Detection of neutrinos from GRBs could be used to test the simultaneity of neutrino and photon arrival to an accuracy of $\sim1{\rm\ s}$, checking the assumption of special relativity that photons and neutrinos have the same limiting speed. These observations would also test the weak equivalence principle, according to which photons and neutrinos should suffer the same time delay as they pass through a gravitational potential. With $1{\rm\ s}$ accuracy, a burst at $1{\rm\ Gpc}$ would reveal a fractional difference in limiting speed of $10^{-17}$, and a fractional difference in gravitational time delay of order $10^{-6}$ (considering the Galactic potential alone). Previous applications of these ideas to supernova 1987A (see \cite{jnb_book} for review), yielded much weaker upper limits: of order $10^{-8}$ and $10^{-2}$ respectively.

\section{Phenomenological considerations}
\label{sec:phenomenology}

\subsection{The luminosity constraint for UHECR sources and its implications for high energy neutrino telescopes}
\label{sec:luminosity}

\begin{figure}[htbp]
\epsfxsize=10cm
\centerline{\epsfbox{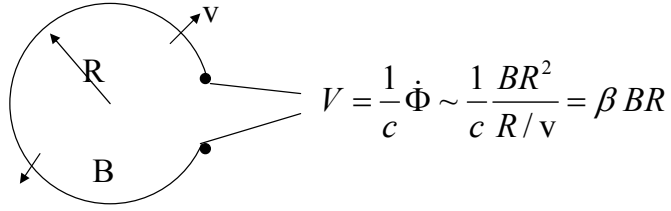}}
\caption{Potential drop generated by an outflow of magnetized plasma.}
\label{fig:acceleration}
\end{figure}

The essence of the challenge of accelerating particles to $>10^{19}$~eV can be understood using the following simple arguments. Consider an astrophysical source driving a flow of magnetized plasma, with characteristic magnetic field strength $B$ and velocity $v$. Imagine now a conducting wire encircling the source at radius R, as illustrated in fig.~\ref{fig:acceleration}. The potential generated by the moving plasma is given by the time derivative of the magnetic flux $\Phi$ and is therefore given by $V=\beta B R$ where $\beta=v/c$. A proton which is allowed to be accelerated by this potential drop would reach energy
$E_p\sim\beta eB R$. The situation is somewhat more complicate in the case of a relativistic outflow, where $\Gamma\equiv(1-\beta^2)^{-1/2}\gg1$. In this case, the proton is allowed to be accelerated only over a fraction of the radius $R$, comparable to $R/\Gamma$. To see this, one must realize that as the plasma expands, its magnetic field decreases, so the time available for acceleration corresponds to the time of expansion from $R$ to, say, $2R$. In the observer frame this time is $R/c$, while in the plasma rest frame it is $R/\Gamma c$. Thus, a proton moving with the magnetized plasma can be accelerated over a transverse distance $\sim R/\Gamma$. This sets a lower limit to the product of the magnetic field and source size, which is required to allow acceleration to $E_p$, $BR>\Gamma E_p/e\beta$. This constraint also sets a lower limit to the rate $L$ at which energy should be generated by the source. The magnetic field carries with it an energy density $B^2/8\pi$, and the flow therefore carries with it an energy flux $>vB^2/8\pi$ (some energy is carried also as plasma kinetic energy), which implies $L>vR^2B^2$ and therefore
\begin{equation}\label{eq:L}
  L>\frac{\Gamma^2}{\beta}\left(\frac{E_p}{e}\right)^2c
  =10^{45.5}\frac{\Gamma^2}{\beta}\left(\frac{E_p}{10^{20}\rm eV}\right)^2{\rm erg/s}.
\end{equation}
Only two types of sources are known to satisfy this requirement. The brightest steady sources are active galactic nuclei (AGN). For them $\Gamma$ is typically between 3 and 10, implying $L>10^{47}{\rm erg/s}$, which may be satisfied by the brightest AGN \cite{Lovelace}. The brightest transient sources are GRBs. For these sources $\Gamma\simeq10^{2.5}$ implying $L>10^{50.5}{\rm erg/s}$, which is generally satisfied since the typical observed MeV-photon luminosity of these sources is $L_\gamma\sim10^{52}{\rm erg/s}$ (It was recognized early on, e.g.  \cite{Hillas} and references therein, that while highly magnetized neutron stars may also satisfy the constraint $BR>\Gamma E_p/e\beta$, it is hard to utilize the potential drop in their electro-magnetic winds for acceleration to ultra-high energy; see, e.g., ref.~\cite{Arons} for a recent discussion).

It is instructive to compare eq.~\ref{eq:L} to the constraints imposed on a neutrino source by requiring it to be detectable by a km-scale neutrino telescope. Consider the minimum flux of a source that can be detected by a neutrino telescope with effective area $A$ (in the plane perpendicular to the source direction) and exposure time $T$. The probability that a muon produced by the interaction of a muon neutrino with a nucleon will cross the detector is given by the ratio of the muon and neutrino mean free paths (Note, that since the mean free path of muons of energy $> 0.3$~TeV is of order 1 km, a detector with effective cross-sectional area $A$ corresponds to a detector with an effective volume $\sim A\times1{\rm  km}$). For water and ice, this probability is $P_{\nu\mu}\approx10^{-4}(E_\nu/100{\rm TeV})^\alpha$, with  $\alpha=1$ for $E_\nu< 100$~TeV and $\alpha=0.5$ for $E_\nu> 100$~TeV (e.g. \cite{GHS}). Thus, the neutrino flux required for the detection of $N$ events is 
\begin{equation}\label{eq:f_nu}
    f_\nu\approx5\times10^{-12}N\left(\frac{E_\nu}{100{\rm TeV}}\right)^{1-\alpha}\left(\frac{AT}{\rm km^2yr}\right)^{-1}{\rm erg/cm^2s}.
\end{equation}
A lower limit to the source flux is also set by the requirement that the signal would exceed the background produced by atmospheric neutrinos. The flux of atmospheric neutrinos, averaged over zenith angle, is approximately given by $\Phi_\nu^A\approx5\times10^{-8}(E_\nu/100{\rm TeV})^{-\beta}{\rm GeV/cm^2s\,sr}$, with $\beta= 1.7$ for $E_\nu< 100$~TeV and $\beta= 2.0$ for $E_\nu> 100$~TeV. For a neutrino detector with angular resolution $\Delta\theta$, the source flux for which the signal constitutes a $5\sigma$ detection over the atmospheric background flux is
\begin{equation}\label{eq:f_nu_A}
    f_\nu\approx3\times10^{-12}\left(\frac{E_\nu}{100{\rm TeV}}\right)^{-\beta/2}\left(\frac{\Delta\theta}{1\rm deg}\right)\left(\frac{AT}{\rm km^2yr}\right)^{-1/2}{\rm erg/cm^2s}.
\end{equation}
Comparing eq.~\ref{eq:f_nu} and eq.~\ref{eq:f_nu_A}, we find that for km-scale detectors, the atmospheric neutrino background poses a less stringent constraint on source flux than the requirement of a detectable signal, except at low energies $\ll100$~TeV.

The luminosity of a cosmologically distant source which corresponds to the flux of eq.~\ref{eq:f_nu} is
\begin{equation}\label{eq:L_nu}
    L_\nu\approx10^{46}Nd_{L,28}^2\left(\frac{E_\nu}{100{\rm TeV}}\right)^{1-\alpha}\left(\frac{AT}{\rm km^2yr}\right)^{-1}{\rm erg/s}.
\end{equation}
Here, $d_L=10^{28}d_{L,28}$~cm is the luminosity distance and $d_{L,28}\sim1$ for sources at redshift $z=1$. The neutrino luminosity is, of course, a lower limit to the total energy output rate from the source. The lower limit set by eq.~\ref{eq:L_nu} to the luminosity of a steady source that may be detectable at $\sim10^{1\pm1}$~TeV neutrino energies by a km-scale neutrino detector, $L>10^{46}{\rm erg/s}$, is similar to the constraint of eq.~\ref{eq:L} imposed by requiring proton acceleration to $\sim10^{20}$~eV. As mentioned following eq.~\ref{eq:L}, the only steady sources bright enough to possibly satisfy these constraints are AGN. Eqs.~\ref{eq:L} and~\ref{eq:L_nu} also imply similar luminosity constraints on GRBs. For a typical GRB duration of $10^{1.5}$~s, eq.~\ref{eq:L_nu} implies $L>10^{52}$~erg/s, similar to the constraint imposed by eq.~\ref{eq:L} for the typical Lorentz factor appropriate for these sources, $\Gamma\simeq10^{2.5}$.  

The phenomenological arguments presented above imply that the luminosity that must be produced by a source of UHECRs, eq.~\ref{eq:L}, may allow such a source to be detectable by a km-scale neutrino detector at $\sim10^{1\pm1}$~TeV neutrino energy, see eq.~\ref{eq:L_nu}. Eq.~\ref{eq:L_nu} also implies that the detection of sources at $\gg10^2$~TeV neutrino energy would require detectors with effective volume $\gg1{\rm km}^3$. The most plausible sources of UHECRs, GRBs and AGN, are therefore also the most likely to be detectable neutrino sources. Such detection would be possible only if the neutrino luminosity of the source constitutes a significant fraction of the total source luminosity. For GRBs, this issue is discussed in some more detail in \S\ref{sec:grb}, within the context of current GRB models.

\subsection{The Waxman-Bahcall bound}
\label{sec:WB}

Cosmic-ray observations suggest that the cosmic-ray flux is dominated above $10^{19}$~eV by extra-Galactic sources of protons, and at lower energy by Galactic heavy nuclei sources. Under the assumption that the UHECRs are extra-Galactic protons, the observed flux of UHECRs allows to determine the rate (per unit time and volume) at which high energy protons are produced \cite{W95b}.
\begin{figure}
\epsfxsize=8cm
\centerline{\epsfbox{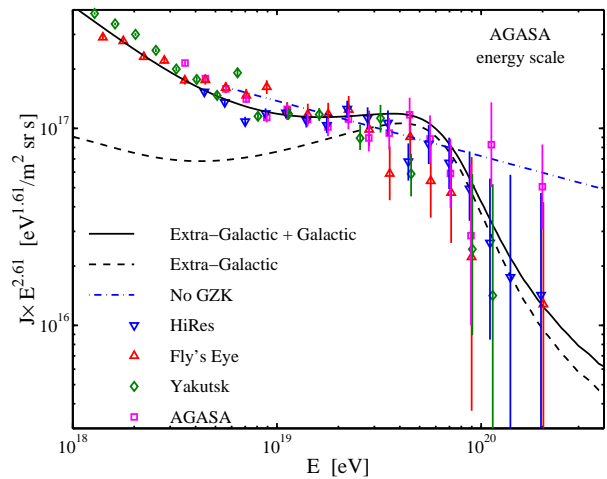}}
\caption{The solid curve shows the energy spectrum derived from the two-component model discussed in \S~2.2 (with $\phi(z)\propto (1+z)^3$ up to $z=2$, following the evolution of star formation rate). The dashed curve shows the extra-Galactic component contribution. The "No GZK" curve is an extrapolation of the $E^{-2.75}$ energy spectrum derived for the energy range of $6\times 10^{18}$~eV to $4\times 10^{19}$~eV \cite{CR_data_rev}. Data taken from \cite{Yakutsk,fly,AGASA,HiRes} (AGASA's energy scale was chosen).}
\label{fig:fig_agasa_norm}
\end{figure}
Figure ~\ref{fig:fig_agasa_norm}, adapted from \cite{BW03}, presents a comparison of available UHECR data with the predictions of a model, where extra-galactic protons in the energy range $E_p\le10^{21}$~eV are produced by cosmologically-distributed sources at a rate and spectrum given by
\begin{equation}
E_p^2\frac{d\dot{N}_p}{dE_p}=0.65\times 10^{44} {\rm erg~Mpc^{-3}~yr^{-1}}\phi(z). \label{eq:energyrate}
\end{equation}
Here, $\phi(z)$ accounts for redshift evolution and $\phi(z=0)=1$. The spectrum above $10^{19}$~eV is only weakly dependent on $\phi(z)$ since proton energy loss limits their propagation distance. For the heavy nuclei component dominating at lower, $<10^{19}$~eV, energy the Fly's Eye experimental fit~\cite{fly}, $dN/dE\propto E^{-3.50}$, was used. The power-law spectrum of accelerated particles, $dN/dE\propto E^{-2}$, has been observed for both non-relativistic and relativistic shocks, and is believed to to be due to Fermi acceleration in collisionless shocks~\cite{Fermi} (although a first principles understanding of the process is not yet available). 

Figure~\ref{fig:fig_agasa_norm} demonstrates that model predictions are in good agreement with the data of all experiments in the energy range $10^{19}$~eV to $10^{20}$~eV. The uncertainty in the derived energy production rate, eq.~\ref{eq:energyrate}, due to systematic uncertainties in the absolute energy calibration of the experiments is $\approx 20\%$ \cite{BW03}. As explained in detail in \cite{BW03}, the various experiments are consistent with each other when systematic errors in the absolute energy scale of the events are taken into account. The relative systematic shifts in absolute energy calibration between Fly' Eye and the other experiments, \{-11\%, +7.5\%, -19\%\} for \{AGASA, HiRes, Yakutsk\}, required to bring into agreement the fluxes measured at $10^{19}$~eV by the different experiments, are well within the published systematic errors. Above $10^{20}$~eV the Fly's Eye, HiRes and Yakutsk experiments are in agreement with each other and with the model, while the AGASA experiment reports a flux higher by a factor $\sim 3$. 

The energy production rate, eq.~\ref{eq:energyrate}, sets an upper bound to the neutrino intensity produced by sources which, like GRBs and AGN jets, are optically thin for high-energy nucleons to $p\gamma$ and $pp(n)$ interactions. For sources of this type, the energy generation rate of neutrinos can not exceed the energy generation rate implied by assuming that all the energy injected as high-energy protons is converted to pions (via $p\gamma$ and $pp(n)$ interactions). The resulting upper bound (for muon and anti-muon neutrinos, neglecting mixing) is \cite{WBbound}
\begin{equation}
E_\nu^2\Phi_\nu<2\times10^{-8}\xi_z\left[\frac{(E_p^2d\dot{N}_p/dE_p)_{z=0}}{10^{44}{\rm erg/Mpc^3yr}}\right]{\rm GeV\,cm}^{-2}{\rm s}^{-1}{\rm sr}^{-1}.
\label{eq:WB}
\end{equation}
$\xi_z$ is (a dimensionless parameter) of order unity, which depends on the redshift evolution of $E_p^2d\dot{N}_p/dE_p$ (see eq.~\ref{eq:energyrate}). In order to obtain a conservative upper bound, we adopt $(E_p^2d\dot{N}_p/dE_p)_{z=0}=10^{44}{\rm erg/Mpc^3yr}$ and a rapid redshift evolution, $\Phi(z)=(1+z)^3$ up to $z=2$, following the evolution of star formation rate. This evolution yields $\xi_z\approx3$. 

\begin{figure}[htbp]
\epsfxsize=8cm
\centerline{\epsfbox{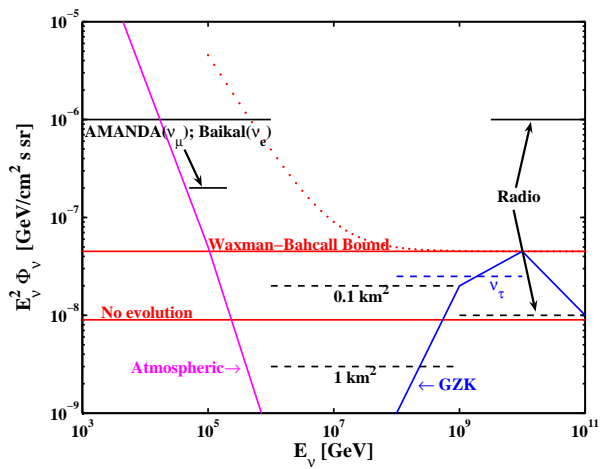}}
\caption{The upper bound imposed by UHECR observations on the extra-Galactic high energy muon neutrino intensity (lower-curve: no evolution of the energy production rate, upper curve: assuming evolution following star formation rate), compared with the atmospheric neutrino background and with the experimental upper bounds (solid lines) of optical Cerenkov experiments, BAIKAL \cite{Baikal} and AMANDA \cite{amanda_bound,Nu_telescopes}, and of coherent Cerenkov radio experiments (RICE \cite{RICE}, GLUE \cite{GLUE}; see also \cite{ANITA}). The dotted curve is the maximum contribution due to possible extra-Galactic component of lower-energy, $<10^{17}$~eV, protons, that may be "hidden under"  the Galactic heavy nuclei flux \cite{WBbound}. The curve labelled "GZK" shows the intensity due to interaction with micro-wave background photons. Dashed curves show the expected sensitivity of 0.1~Gton (AMANDA, ANTARES, NESTOR) and 1~Gton (IceCube, NEMO) optical Cerenkov detectors \cite{Nu_telescopes}, of the coherent radio Cerenkov (balloon) experiment ANITA \cite{ANITA} and of the Auger air-shower detector (sensitivity to $\nu_\tau$) \cite{Saltzberg}. Space air-shower detectors (OWL-AIRWATCH) may also achieve the sensitivity required to detect fluxes lower than the WB bound at energies $>10^{18}$~eV \cite{Saltzberg}.}
\label{fig:WBbound}
\end{figure}

The WB upper bound is compared in fig.~\ref{fig:WBbound} with current experimental limits, and with the expected sensitivity of planned neutrino telescopes. 
The figure indicates that km-scale (i.e. giga-ton) neutrino telescopes are needed to detect the expected extra-Galactic flux in the energy range of $\sim1$~TeV to $\sim1$~PeV, and that much larger effective volume is required to detect the flux at higher energy.

\subsection{The WB bound and AGN jet model predictions}
\label{sec:AGN}

\begin{figure}[htbp]
\epsfxsize=8cm
\centerline{\epsfbox{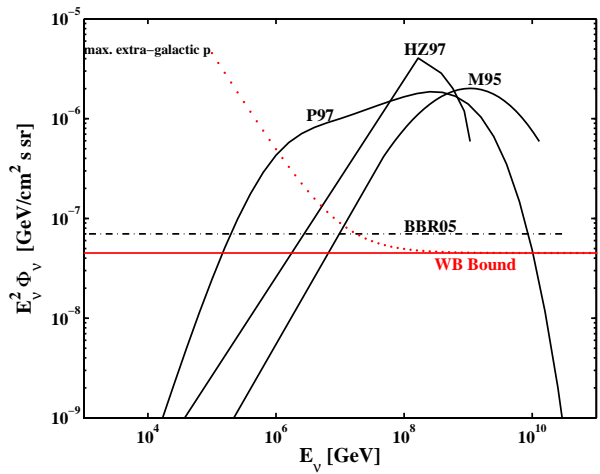}}
\caption{The WB bound and various model predictions for the diffuse neutrino intensity produced by AGN jets. Early models (M95=Mannheim 1995, HZ97=Halzen \& Zas 1997, P97=Protheroe 1997 \cite{early_agn}) predicted an intensity well above the WB bound, and are therefore inconsistent with UHECR observations. More recent models (BBR05= Becker, Biermann \& Rhode \cite{late_agn}) predict fluxes close to the WB bound.}
\label{fig:agn}
\end{figure}

Figure~\ref{fig:agn} presents a comparison of the WB bound with various model predictions for the diffuse neutrino intensity produced by AGN jets. Early models \cite{early_agn} predicted an intensity well above the WB bound, and are therefore inconsistent with UHECR observations. More recent models \cite{late_agn}, with more detailed consideration of plasma conditions in the jets, predict fluxes close to the WB bound. 

\subsection{GZK neutrinos}
\label{sec:GZK}

Protons of energy exceeding the threshold for pion production, $\sim5\times10^{19}$~eV, lose most of their energy over a time short compared to the age of the universe. $>10^{20}$~eV protons, for example, lose most of their energy over $<3\times10^8$yr. Assuming that high-energy protons are produced by extra-Galactic sources at the rate implied by UHECR observations (eq.~\ref{eq:energyrate}), the proton energy loss to pion production produces a neutrino intensity similar to the WB-bound. The expected GZK neutrino intensity is schematically shown in fig.~\ref{fig:WBbound}. Since most of the pions are produced in interactions with photons of energy corresponding to the $\Delta$-resonance, each of the resulting neutrinos carry approximately 5\% of the proton energy. The neutrino background is therefore close to the WB bound above $\sim5\times10^{18}$~eV, where neutrinos are produced by $\sim10^{20}$~eV protons. The intensity at lower energies is lower, since protons of lower energy do not lose all their energy over the age of the universe. The GZK intensity in figure~\ref{fig:WBbound} decreases at the highest energies since it was assumed that the maximum energy of protons produced by UHECR sources is $10^{21}$~eV. The results of detailed calculations of the expected GZK neutrino intensity \cite{ESS01} are in agreement with the qualitative analysis presented above.

The detection of GZK neutrinos will be a milestone in neutrino astronomy. Most important, neutrino detectors with sensitivity better than the WB-bound at energies $>10^{18}$~eV will test the hypothesis that the UHECR are protons (possibly somewhat heavier nuclei) of extra-Galactic origin. Measurements of the flux and spectrum would constrain the redshift evolution of the sources. The GZK neutrino intensity presented in fig.~\ref{fig:WBbound} is obtained for rapidly evolving sources. For non-evolving sources the intensity is lower by a factor of $\approx5$.

The uncertainty in the experimental determination of the UHECR flux at energies $>10^{20}$~eV lead many authors to speculate that there may be a new source of ultra-high energy cosmic-rays and neutrinos beyond $10^{20}$~eV, producing a neutrino intensity higher than the WB bound at these neutrino energies. Most of these models involve "new physics": Decay of super-massive dark matter particles and topological defects (e.g.~\cite{BhattacharjeeSigl}) have been proposed to produce ultra-high energy neutrinos with small associated cosmic ray proton flux, thus possibly producing a flux of neutrinos exceeding the WB-bound. In our view, it is reasonable to extend the WB limit beyond $10^{20}$~eV by simply extrapolating the horizontal line in fig.~\ref{fig:WBbound} to higher energies. The validity of this extrapolation will be tested by future measurements of the spectrum of ultra-high energy cosmic rays and neutrinos.

\section{Gamma-ray bursts (GRBs): An illustrative example}
\label{sec:grb}

\subsection{The fireball model and the association of GRBs and UHECRs}
\label{sec:fireball}

Gamma-ray bursts (GRBs) are short, typically tens of seconds long,
flashes of gamma-rays, carrying most of their energy in $>1$~MeV photons. The detection in the past few years of "afterglows," delayed X-ray, optical and radio emission from GRB sources, proved
that the sources lie at cosmological distances, and provided strong support for the following scenario of GRB production \cite{GRBrev,Mrev}. The energy source is believed to be rapid mass accretion on a newly formed solar-mass black hole. Recent observations suggest that the formation of the central compact object is associated with type
Ib/c supernovae \cite{GRBSNe}. The energy release drives an ultra-relativistic, $\Gamma\sim10^{2.5}$, plasma outflow. The emission of $\gamma$-rays is assumed to be due to internal collisionless shocks within the relativistic wind, the "fireball," which occur at a large distance from the central black-hole due to variability in the wind emitted from the central "engine." It is commonly assumed that electrons are accelerated to high
energy within the collisionless shocks, and that synchrotron emission from these shock accelerated
electrons produces the observed $\gamma$-rays. If protons are present in the wind, as assumed in the fireball model, they would also be accelerated to high energy in the region were electrons are accelerated.

GRBs were suggested to be UHECR sources in \cite{W95,VMU}. The GRB-UHECR association was based in \cite{W95} on two major arguments (see \cite{Wrev}) for a pedagogical review). First, the constraints imposed on the relativistic wind by requiring that it would produce the observed $\gamma$-rays were shown to be remarkably similar to those imposed by the
requirement that the wind would allow proton acceleration (in the internal collisionless shocks) to $10^{20}$~eV. Independent physical arguments lead, in both cases, to similar lower limits on the expansion Lorentz factor, $\Gamma\ge300$, and on the ratio of magnetic field and electron energy densities, $u_B/u_e\ge0.1$. Second, the rate (per unit volume) at which energy is generated by GRBs in $\gamma$-rays was shown to be similar to the rate at which
energy should be generated in high energy protons in order to account for the observed UHECR flux.
Both arguments have been strengthened by more recent GRB and UHECR observations \cite{W04}.

\subsection{100~TeV neutrinos}
\label{sec:1e14}

Protons accelerated in the fireball internal shocks lose energy through photo-meson interaction with fireball photons. The decay of charged pions produced in this interaction results in the
production of high energy neutrinos. The key relation is between the observed photon energy, $E_\gamma$, and the accelerated proton's energy, $E_p$, at the threshold of the
$\Delta$-resonance. In the observer frame,
\begin{equation}
E_\gamma \,E_{p} = 0.2 \, {\rm GeV^2} \, \Gamma^2\,.
\label{eq:keyrelation}
\end{equation}
For $\Gamma\approx10^{2.5}$ and $E_\gamma=1$~MeV, we see that
characteristic proton energies $\sim 10^{16}$~eV are required to
produce pions. Since neutrinos produced by pion decay typically
carry $5\%$ of the proton energy, production of $\sim 10^{14}$~eV
neutrinos is expected \cite{WnB97}.

The fraction of energy lost by protons to pions, $f_\pi$, is $f_\pi\approx0.2$ \cite{WnB97,Wrev}. Assuming that GRBs generate the observed UHECRs, the expected
GRB muon-neutrino flux may be estimated using eq.~\ref{eq:WB} \cite{WnB97,WBbound}, 
\begin{equation}
E_\nu^2\Phi_{\nu}\approx
0.3\times10^{-8}{f_\pi\over0.2}{\rm GeV\,cm}^{-2}{\rm s}^{-1}{\rm
sr}^{-1}. \label{eq:JGRB}
\end{equation}
This neutrino spectrum extends to $\sim10^{16}$~eV, and is suppressed at higher energy due to energy loss of pions and muons \cite{WnB97,RnM98,WBbound}. Eq.~\ref{eq:JGRB} implies a detection rate of $\sim20$ neutrino-induced muon events per year (over $4\pi$~sr) in a cubic-km detector. Since GRB neutrino events are correlated both in time and in direction with gamma-rays, their detection is practically background free.

\subsection{TeV neutrinos}
\label{sec:TeV}

The 100~TeV neutrinos discussed in the previous section are produced at the same region where GRB $\gamma$-rays are produced. Their production is a generic prediction of the fireball model. It is a direct consequence of the assumptions that energy is carried from the underlying engine as kinetic energy of protons and that $\gamma$-rays are produced by synchrotron emission of shock accelerated particles. Neutrinos may be produced also in other stages of fireball evolution, at energies different than 100~TeV. The production of these neutrinos is dependent on additional model assumptions. We discuss below some examples related to the GRB progenitor. For a more detailed discussion see \cite{Mrev,Wrev} and references therein.

The most widely discussed progenitor scenarios for long-duration GRBs involve core collapse of massive stars. In these "collapsar" models, a relativistic jet breaks through the stellar envelope to produce a GRB. For extended or slowly rotating stars, the jet may be unable to break through the envelope. Both penetrating (GRB producing) and "choked" jets can produce a burst of  $\sim5$~TeV neutrinos by interaction of accelerated protons with jet photons, while the jet propagates in the envelope \cite{choked}. The estimated event rates may exceed $\sim10^2$ events per yr in a km-scale detector, depending on the ratio of non-visible to visible fireballs. A clear detection of non-visible GRBs with neutrinos may be difficult due to the low energy resolution for muon-neutrino events, unless the associated supernova photons are detected. In the two-step "supranova" model, interaction of the GRB blast wave with the supernova shell can lead to detectable neutrino emission, either through nuclear collisions with the dense supernova shell or through interaction with the intense supernova and backscattered radiation field \cite{supranova}.

\end{document}